\documentclass[12pt,preprint2]{aastex}

\shorttitle{Discovery of Stellar Companions to Two Exoplanet Host Stars}
\shortauthors{Roberts et al.}

\begin{document}


\title{Know the Star, Know the Planet. III. Discovery of Late-Type Companions to Two Exoplanet Host Stars}

\author{
Lewis C. Roberts, Jr.\altaffilmark{1},  
Andrei Tokovinin\altaffilmark{2}, 
Brian D. Mason\altaffilmark{3},
Reed L. Riddle\altaffilmark{4}, 
William I. Hartkopf\altaffilmark{3},
Nicholas M. Law\altaffilmark{5}, and 
Christoph Baranec\altaffilmark{6}
} 

\altaffiltext{1}{Jet Propulsion Laboratory, California Institute of Technology, 4800 Oak Grove Drive, Pasadena CA 91109, USA}
\altaffiltext{2}{Cerro Tololo Inter-American Observatory, Casilla 603, La Serena, Chile}
\altaffiltext{3}{U.S. Naval Observatory, 3450 Massachusetts Avenue, NW, Washington, DC 20392-5420, USA}
\altaffiltext{4}{Division of Physics, Mathematics, and Astronomy, California Institute of Technology, Pasadena, CA 91125, USA}
\altaffiltext{5}{Department of Physics and Astronomy, University of North Carolina at Chapel Hill, Chapel Hill, NC 27599-3255, USA}
\altaffiltext{6}{Institute for Astronomy, University of Hawai$\textquoteleft$i at M\={a}noa, Hilo, HI 96720-2700, USA}

\email{lewis.c.roberts@jpl.nasa.gov}
 

\begin{abstract}

We discuss two multiple star systems that host known exoplanets: HD 2638 and 30 Ari B.   Adaptive optics imagery revealed an additional stellar companion to both stars. We collected  multi-epoch images of the systems with Robo-AO and the PALM-3000 adaptive optics systems at Palomar Observatory and provide relative photometry and astrometry.  The astrometry indicates that the companions share common proper motion with their respective primaries.  Both  of the new companions have projected separations less than 30 AU from the exoplanet host star. Using the projected separations to compute orbital periods of the new stellar companions, HD 2638 has a period of 130 yrs and 30 Ari B has a period of 80 years.  Previous studies have shown that the true period is most likely within  a factor of three of these estimated values.  The additional component to the 30 Ari   makes it the second confirmed quadruple system known to host an exoplanet. HD 2638 hosts a hot Jupiter and the discovery of a new companion strengthens the connection between hot Jupiters and binary stars.    We place the systems on a color-magnitude diagram and derive masses for the companions which turn out to be roughly 0.5 solar mass stars.

\end{abstract}

\keywords{binaries: visual - instrumentation: adaptive optics - stars: individual (HD 2638, 30 Ari B) stars: solar-type}
  

\section{INTRODUCTION}
 
Since the discovery of the first exoplanet system in the 1990s, planetary systems have been found in an increasingly varied array of architectures.   With this variety has come an increasing number of models and theories to explain  these systems. It has also been realized that stellar companions play a key role in the evolution of the planetary dynamics for some systems.   There are several proposed ways in which binary stars can influence the orbital properties of exoplanets. 

One of these is Kozai migration \citep{wu2003, wu2007, fabrycky2007, takeda2008} where the mutual torques between the binary and the planet transfer angular momentum, causing the orbit of the planet to become more eccentric. Eventually the eccentricity becomes high enough that the planet approaches the star, which then tidally circularizes the orbit. Not all exoplanets in binary systems undergo Kozai migration, as it depends on the mutual inclination of the orbits as well as the architecture of the planetary system. \citet{naoz2012} found that it can account for about 30\% of the observed hot Jupiter planets, which matches well with the projected spin-orbit angle distribution of hot Jupiters.  

The stellar companion can also interact with the protoplanetary disk. \citet{rafikov2013} proposed that the stellar companion effects the protoplanetary disk by slowing the planetesimals in the disk and allowing the formation of planets.  This requires a relatively massive disk, but can produce systems such as $\gamma$ Cep and $\alpha$ Cen \citep{rafikov2013}.  

\citet{kley2008} suggest an alternative method in which an eccentric stellar companion leads to a strong periodic disturbance in the disk whenever the companion is at periastron.  This leads to an eccentric disk that changes over time due to the interactions with the stellar companion.  This can produce massive eccentric planets, depending on the initial position of the planetesimal.   Their modeling did not consider the initial creation of the planetesimals.

Circumbinary planets have also been discovered, where the planet orbits both stars in a short-period orbit, on the order of tens of days \citep{lee2009, doyle2011}. In these systems  the planet(s) and the stellar companion are all co-planar implying that they formed from flat circumbinary disks. Since their discovery there is considerable effort to model the formation and dynamics of these systems \citep{kley2014, armstrong2014}. 

In the case of multiple stellar systems, coplanar systems are rare. Of the 2413 systems cataloged in the {\it Sixth Catalog of Orbits of Visual Binary Stars} \citep{orb6} 48 systems have orbit solutions with all seven Campbell elements for both inner and outer systems in hierarchical arrangements. Of these, only four have the possibility of being coplanar according to the precepts of \citet{fekel81}.

The various models of exoplanet formation in binary systems can only be fully tested when there are sufficient numbers of known systems, especially systems with relatively short binary periods.  There have been a number of surveys to measure the duplicity rates among the exoplanet hosts \citep{patience2002, eggenberger2007, chauvin2011,  mason2011, roberts2011, ginski2012}. These surveys have revealed several interesting features. 
 
The frequencies of exoplanets among single stars and components of wide binaries (semi-major axis larger than ~100 AU) are indistinguishable \citep{raghavan2006, bonavita2007}.  Also, the population of exoplanets in wide binaries  is essentially the same as the population around single stars, indicating that wider binaries have minimal impact on dynamics of the exoplanets \citep{bonavita2007}.  The population of exoplanets in binaries with semi-major axis smaller than 100 AU is statistically different than those orbiting single stars \citep{zucker2002,bonavita2007}.  There are fewer exoplanets in tight stellar binary systems, but the exoplanets  tend to be more massive \citep{desidera2007,duchene2010}.   No planet has been detected in stellar binaries that have a separation of less than 10 AU \citep{roell2012}. That same paper also determined that multi-exoplanet systems have only been detected in binary stars with a  projected separation larger than 100AU.

Known close visual binaries are traditionally excluded from radial-velocity (RV) exoplanet programs because the presence of a visual companion degrades the RV precision. This intrinsic bias complicates statistical inferences about exoplanets in binaries. Moreover, a faint visual companion that is itself a close (spectroscopic) binary pair can produce periodic low-amplitude RV modulation in the combined light that can be mistaken for an exoplanet. False positive exoplanet detections caused by  unrecognized hierarchical multiplicity of their hosts may reach 1--2 percent \citep{tokovinin2014b} of the total exoplanet sample. This is yet another reason to observe exo-hosts with high angular resolution and deep dynamical range. 

In addition to dynamical interactions, exoplanets and their host stars share a common origin, and determining the properties of that local environment for the star leads to information about the exoplanet. Examples of this include metallicity correlations (e.g. 16 Cyg AB, \citealt{schuler2011}), stellar rotation, and activity.

This paper details two multiple star systems that host known exoplanets: HD 2638 and 30 Ari B. Additional stellar components were found in each of these systems by \citet{riddle2015}. This paper discusses the systems and those newly discovered components systems in depth.  The observations of the systems are covered in Section \ref{observations}. Section \ref{hd2638} discusses the HD 2638 system and Section \ref{30AriB} discusses the 30 Ari system.  Our results are summarized in Section \ref{summary}. 
 

\section{OBSERVATIONS}\label{observations}

\subsection{ROBO-AO}
The companions to HD 2638 and 30 Ari B were first detected using the Robo-AO system as part of a survey of nearby binaries with solar type components \citep{riddle2015}. We have copied the astrometry and photometry from that paper in order to do a more in-depth analysis of these two stars. For those unfamiliar with Robo-AO, we provide a brief introduction to the system.  Robo-AO is a  robotic laser guide star adaptive optics (AO) system \citep{baranec2013,baranec2014}  that operates automatically in order to carry out high efficiency observing programs. Robo-AO is currently deployed on the Palomar Observatory 60-inch telescope.  Robo-AO uses Rayleigh scattering from a UV laser focused at 10 km from the telescope as the wavefront reference, and generates diffraction limited images with Strehl ratios of 4-26\% in the $i$ filter. The AO system corrects the high order wavefront aberrations with automated software that operates at a rate of 1.2 kHz, sharpening the instantaneous point spread  function across the science camera field of view (FOV). A bright star within the FOV is still required to correct the tip-tilt motion, which is corrected by the automated data processing software.  
 
The Robo-AO science camera is an electron multiplying CCD (EMCCD) detector with 1024$\times$1024 pixels; the pixel scale is 43.1 mas/pixel, with a total FOV of 44\arcsec$\times$44\arcsec. The EMCCD is operated at 8.6 frames per second in science mode, with an exposure time of 0.115 s each. A total of 516 frames are gathered during each 60 s exposure, which are then combined into a single image for further analysis by the automated data processing software. Both targets were observed with the SDSS \textit{i\'} filter \citep{york2000}, and additionally HD 2638 was observed with the SDSS \textit{r\'} and \textit{z\'} bands. The data were collected during multiple observing runs in 2012 and 2013.  See Sections \ref{hd2638} and \ref{30AriB} for details.   The Robo-AO data are procesed by the  standard pipeline, which is discussed in \citet{riddle2015}.  That paper did not publish error bars for the astrometry measured using speckle processing. It did give mean errors on measurements of control binaries. These companions are fainter than the control binaries, so to be conservative we have set the error bars as 10 mas, which translates to 1\fdg1 in position angle at 0\farcs5 separation. 

\subsection{PALM-3000 Adaptive Optics}

In order to confirm and further characterize the companions, we observed both stars on 2013 September 28 UT with the Palomar Observatory Hale 5 m telescope using the PALM-3000 AO system  and the PHARO near-IR camera.    The PALM-3000 AO system is a natural guide star system using two deformable mirrors \citep{dekany2013}. One corrects low-amplitude high spatial frequency aberrations, while the other corrects the higher-amplitude low spatial frequency aberrations. It is optimized for high contrast observations and routinely produces Strehl ratios  greater than 80\% in \textit{Ks} band.   The PHARO camera uses a HgCdTe HAWAII detector for observations between 1 and 2.5\micron~wavelength \citep{hayward2001}. The camera has multiple filters in two filter wheels.  For HD 2683 we collected 100 frames in both the  \textit{J} and \textit{Ks} filters, while we collected 50 frames of 30 Ari B in the same filters.   After debiasing, flat fielding, bad pixel correction and background subtraction, the frames were coadded.    After the individual frames were calibrated, we created multiple images, by coadding 10 frames into each image. The iterative blind deconvolution  algorithm, \textit{fitstars}, was used to measure the astrometry and photometry of the objects \citep{tenBrummelaar1996,tenBrummelaar2000}.  

Photometric error bars were assigned using the technique described in \citet{roberts2005}.  For the astrometric error analysis, we computed the standard deviation of the astrometric measurements.  Analysis of simulated binaries created using the technique of \citet{roberts2005} showed that this error is almost entirely measurement error.  The other major sources of astrometric error are the errors in the calibration of the plate scale and the position angle offset.  The PHARO  pixel scale has been measured many times with a combination of an astrometric mask and binary star measurements \citep{metchev2006}. The changes between observing runs are small.  On the night these data were acquired, there was not enough time to repeat the lengthly astrometric mask calibration process, but we observed six calibration binaries and updated the plate scale and the orientation angle offset.  This resulted in a plate scale of 0\farcs025$\pm$0\farcs0005 and a position angle offset of 0\fdg7$\pm$0\fdg5.  The final error bar on the astrometry was the root sum square (RSS) of the calibration errors and the scatter in the astrometric measurements.


\section{HD 2638}\label{hd2638}

HD 2638 (HIP 2350 = WDS 00293--0555) is a nearby solar type star with an estimated age of 3 Gyr \citep{moutou2005}. It is paired with  HD 2567 (HIP 2292), another solar type star, in a wide (839\arcsec) common-proper-motion binary. \citet{shaya2011} determined that the pair had a 99\% chance of being true companions through a Bayesian analysis of the measured astrometry.      The  astronomical details of the HD 2638 system are given in Table \ref{tab:objects}.  The measured parallaxes from Hipparcos of the two stars are slightly different. For HD 2638 it is 20.03$\pm$1.49 mas and for HD 2567 it is 17.63$\pm$0.79 mas \citep{vanLeeuwen2007}. If the parallaxes are correct, it would mean that the stars are separated by 6.8 pc, which would mean they are not physically bound. This would disagree with the conclusion of \citet{shaya2011}.  Independent confirmation that A and B are physical comes from their measured RVs being almost identical. +9.60$\pm$0.3 km/s for A \citep{nordstrom2004}, and +9.61$\pm$0.01 for B \citep{latham2002}.  For the distance to the system, we will use the value for HD 2567 since it has the lower error bar.   Unresolved binary stars, such as turns out to be the case with HD 2638 (see below), can cause  errors in parallaxes and proper motions measured by Hipparcos due to  orbital motion in the case of close pairs  \citep{shatskii1998} and due to a cross-talk with the scanning law for wider pairs  \citep{tokovinin2013}. This most likely explains the difference.

\begin{deluxetable}{llllccc} 
\tablewidth{0pt}
\tablecaption{Basic information on two multiple systems \label{tab:objects}}
\tablehead{
\colhead{Object} &
\colhead{WDS} &
\colhead{HD} &
\colhead{HIP} &
\colhead{$\pi_{\rm HIP}$} &
\colhead{$V$} &
\colhead{Spectral} \\
& 
\colhead{(J2000)} &
\colhead{} &
\colhead{} &
\colhead{(mas)} &
\colhead{(mag)} &
\colhead{type}
 }
\startdata
HD 2567 A   & 00293$-$0555 & \phn2567 & \phn2292 & 17.63$\pm$0.79 & 7.76 & G0 \\
HD 2638 B   &              & \phn2638 & \phn2350 & 20.02$\pm$1.49 & 9.37 & G5 \\
30 Ari A    & 02370$+$2439 &    16246 &    12189 & 23.93$\pm$0.59 & 6.48 & F5V \\
30 Ari B    &              &    16232 &    12184 & 24.52$\pm$0.68 & 7.09 & F6V 
\enddata
\end{deluxetable}

Radial velocity (RV) observations detected a hot Jupiter orbiting HD 2638 \citep{moutou2005}. The planet, HD 2638b, has a minimum mass of 0.48$M_{\mathrm{Jup}}$ and an orbital period of  3.4442 $\pm$ 0.0002 d. The orbital semi-major axis is 0.044 AU and the orbit has been tidally circularized.  The planet's equilibrium temperature is on the order of 1100 K (with an assumed albedo of 0.3).  

\begin{figure*}[htb]
    \centering
    \includegraphics[height=6cm]{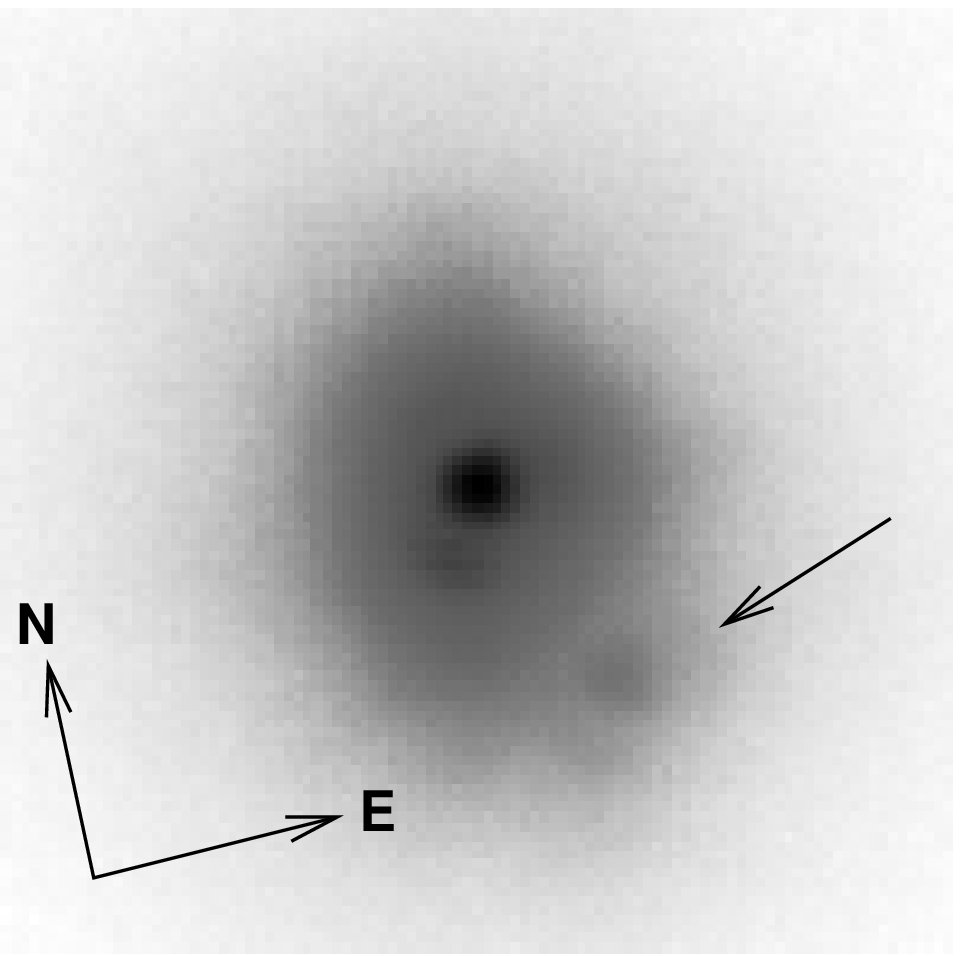}\label{hd2638_a}  
    \includegraphics[height=6cm]{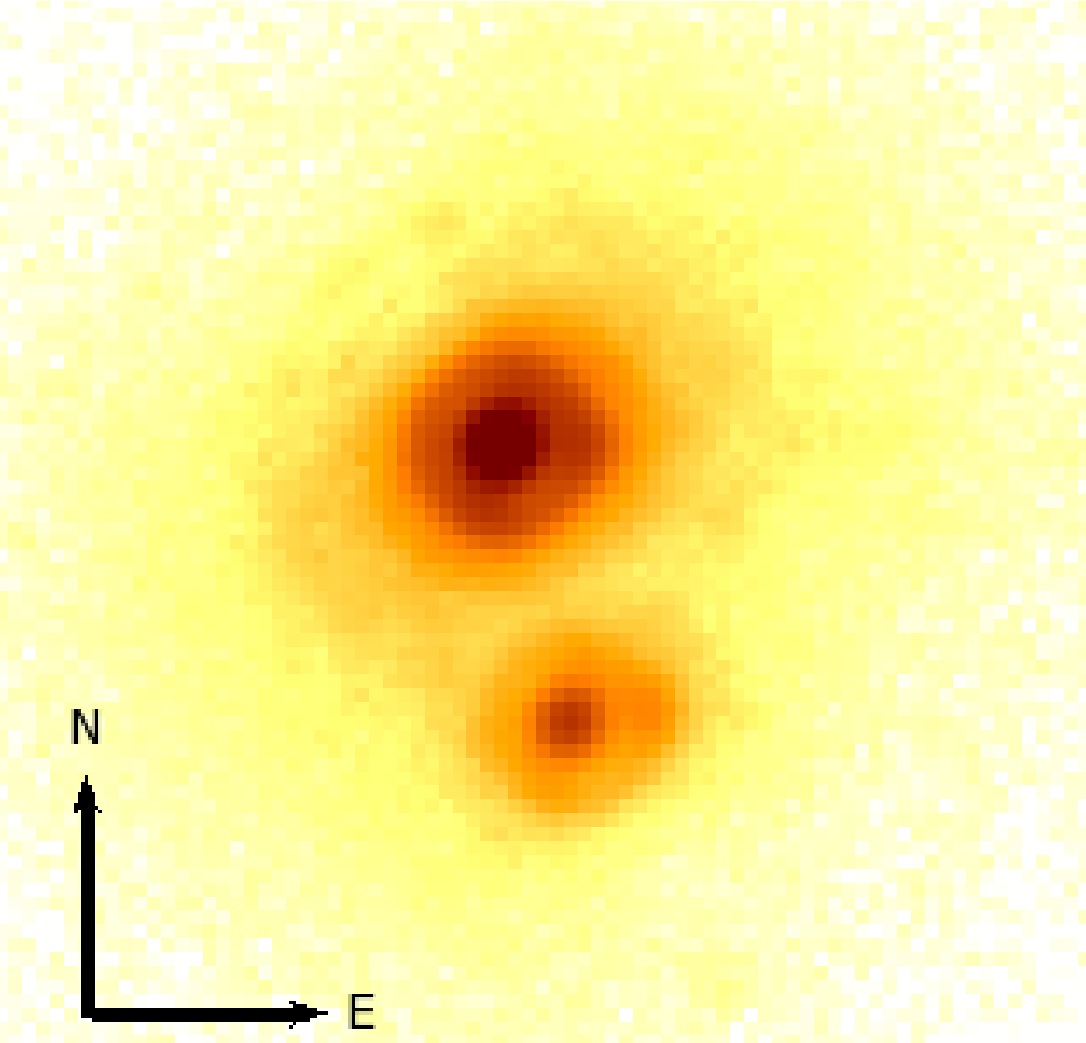}\label{hd2638_b}
    \caption{ On the left is an $i$ image of the HD 2638 BC from Robo-AO.  It is a subimage of the full 44\arcsec~field of view and is $\sim$2\farcs5~across. An arrow points to the location of the newly discovered companion. On the right is a $Ks$ image of the HD 2638 BC acquired with the PALM 3000 AO system. It is a subimage of the full 25\arcsec~field of view and is $\sim$ 2\arcsec~across.
}
    \label{hd2638_image}
\end{figure*}

\citet{riddle2015} imaged a companion to HD 2638 on 2012 September 3 UT.  Over the next 13 months,  two more observations were carried out  with Robo-AO \citet{riddle2015} and another observation was made with the 5 m Hale telescope using  the Palm 3000 AO system and the near-IR PHARO camera. Images of HD 2638 BC are presented in Figure \ref{hd2638_image}.     The measured astrometry and photometry are shown in Table \ref{results_hd2638}. In \citet{riddle2015} there are two measurements from 2012.7621. One has a position angle of 167\fdg3 and the other is 170\fdg0. The later measurement clearly is in error, as it does not  fit the progression of the astrometry. We have chosen to ignore that astrometric point.  
 
\begin{figure*}[htb]
   \centering
   \includegraphics[height=60mm]{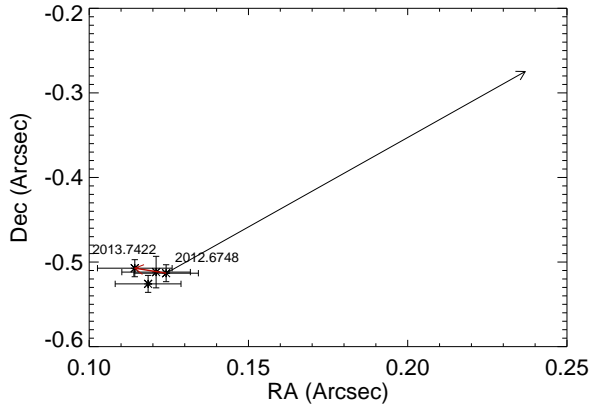}  
   \caption{ The measured astrometry for HD 2638 BC is plotted. The black arrow is the position the companion would have been at our last observation if it was a background star with no proper motion.  The red line shows the actual motion between our first and last observation.  }
   \label{hd2638_motion}
\end{figure*}

HD 2638 has a proper motion of 0.247 $\pm$ 0.0012\arcsec/yr \citep{vanLeeuwen2007}.  We compare this to the relative motion of the companion to the primary of 0.0108 $\pm$ 0.0175\arcsec/yr. It is also in a different direction as shown in Figure \ref{hd2638_motion}.    While the error bar on the relative motion is high compared to the motion, both are much smaller than the proper motion and we conclude that the companion has common proper motion with HD 2638. The measured motion is consistent with orbital motion in that both the position angle and separation are changing in a regular manner within the error bars.  With only four data points points it is impossible to determine if there is a deviation from linear motion, which would be required to confirm orbital motion.  The binary pair is designated HD2638 BC. Normally the pair would be designated HD 2638 Ba,Bb, but that could cause confusion with the planetary companion. 
 
\subsection{DISCUSSION}

The components of the newly discovered binary system can be placed on the color-magnitude diagram (CMD), using the photometry in the combined light from the literature and the differential photometry we measured with Robo-AO and PHARO. We also use the the Dartmouth isochrones for 1\,Gyr  and 3\,Gyr age with solar metallicity  \citep{dotter2008}\footnote{ \url{http://stellar.dartmouth.edu/models/webtools.html} }. The isochrones define the relations between various colors and between mass and absolute magnitude.  The combined $J$ and $K$ magnitudes are taken from 2MASS, but the combined SDSS $i$ magnitudes for our targets are not measured directly. We determined $i$ by extrapolating from the known $V$ and $J$ magnitudes, using polynomial relations between the $V-i$, $i-J$, and $V-K$ colors derived from the   Dartmouth isochrones. The combined, differential, and individual magnitudes of the components are listed in Table~\ref{tab:ptm}. 

\begin{figure*}[htb]
   \centering
   \includegraphics[height=100mm]{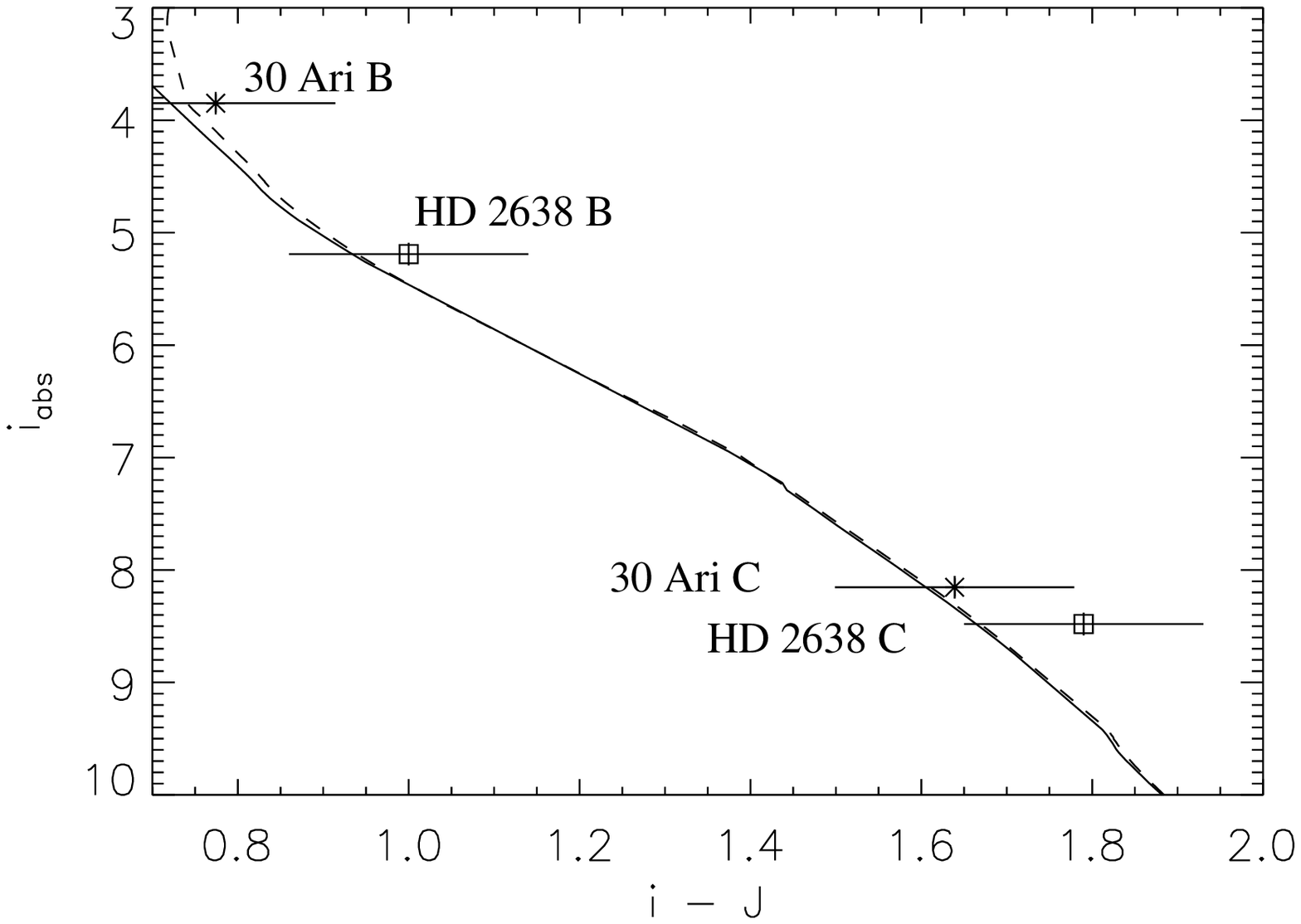}  
   \caption{Position of the binary components in the color-magnitude diagram. The full line shows a 1\,Gyr isochrone (correspondong to 30 Ari B) and the dashed line shows the 3\,Gyr isochrone (corresponding to HD 2638), both with solar metallicity \citep{dotter2008}.    }
   \label{cmd_figure}
\end{figure*}

Figure~\ref{cmd_figure} shows the position of the binary components in the $(i, i -J)$ CMD, together with the 3-Gyr isochrone defining the main sequence. The masses in Table~\ref{tab:ptm} are determined from the isochrone and the absolute magnitudes in the $i$ band.

\begin{deluxetable}{lllcccccc} 
\tablewidth{0pt}
\tablecaption{Relative astrometry and photometry of HD 2638 BC. \label{results_hd2638}}
\tablehead{\colhead{Epoch} & \colhead{$\theta$ ($^{\circ}$)} &\colhead{$\rho$ (\arcsec)}& \colhead{\textit{$\Delta$r}} & \colhead{\textit{$\Delta$i}} & \colhead{\textit{$\Delta$z}} & \colhead{\textit{$\Delta$J}}&\colhead{\textit{$\Delta$Ks}}& \colhead{Instrument} }
\startdata
2012.6748 & 166.4$\pm$1.1 & 0.528$\pm$0.010   & ...  & 3.38 & ...  & ...           & ...           & Robo-AO\\
2012.7621 & 168.7$\pm$1.1 & 0.553$\pm$0.010   & 3.87 & ...  & 3.06 & ...           & ...           & Robo-AO\\
2013.6208 & 166.7$\pm$1.1 & 0.526$\pm$0.019   & ...  & 3.19 & ...  & ...           & ...           & Robo-AO\\
2013.7422 & 167.3$\pm$1.4 & 0.52\phn$\pm$0.01 & ...  & ...  & ...  & 2.50$\pm$0.09 & 1.99$\pm$0.05 & PHARO\\
\enddata
\end{deluxetable}

\begin{deluxetable}{ l cccc} 
\tablewidth{0pt}
\tablecaption{Photometry of the new pairs \label{tab:ptm}}
\tablehead{
\colhead{~} &
\colhead{$i$} &
\colhead{$J$} &
\colhead{$Ks$} &
\colhead{Mass} \\
&
\colhead{(mag)} &
\colhead{(mag)} &
\colhead{(mag)} &
\colhead{ (${\cal M}_\odot$)}
 }
\startdata
\multicolumn{5}{c}{HD 2638 BC, $m-M=3.77$ mag} \\
BC combined   & \phn8.91 & \phn6.08 & 5.82 & \ldots \\
BC $\Delta$m  & \phn3.29 & \phn2.50 & 1.99 & \ldots \\
~~~B          & \phn8.96 & \phn7.96 & 7.47 & 0.87 \\
~~~C          &    12.25 &    10.46 & 9.46 & 0.46 \\
\multicolumn{5}{c}{30 Ari BC, $m-M=3.05$ mag} \\
BC combined   & \phn6.88 & \phn6.08 & 5.82 & \ldots \\
BC $\Delta$m  & \phn4.31 & \phn3.44 & 2.84 & \ldots \\ 
~~~B          & \phn6.90 & \phn6.13 & 5.90 & 1.11 \\
~~~C          &    11.20 & \phn9.57 & 8.74 & 0.50 
\enddata
\end{deluxetable}

We estimate the orbital period, $P$, of the binary from the measured separations, $\rho$, distances, and mass sum ${\cal M}$  using Kepler's Third Law, $\frac{a^3}{P^2} = \cal M$. The median ratio between the projected separation, $d$ $( = \frac{\rho}{\pi_{\rm HIP}})$, and orbital semi-major axis, $a$, is close to one \citep{tokovinin2014a}, with scatter by a factor of two caused by orbital phase, orbit orientation, and eccentricity. The strict lower limit is $a > \frac{d}{2}$. Statistical period estimates using the assumption $a=d$ are denoted as $P^*$. 

The masses of HD~2638 B and C estimated from their absolute $i'$ magnitudes correspond to the spectral types G8V and M1V. The projected separation of 28.5\,AU leads to $P^* \sim 130$\,yr. The true period  likely equals $P^*$ to within a factor of three.  There is however a small probability that the true separation is much larger than the projected separation, in which case the actual period can be substantially longer than $P^*$.

The projected separation of HD~2638 B and C is quite close.  The stellar companion has most likely had a significant impact on the dynamical evolution of the planet.  It has long been thought that hot Jupiters formed beyond the snow line and migrated to their current locations. There are a number of theories on what caused that migration.  One suggested mechanism is that interaction between the protoplanetary disk and the planet causes the planet to migrate inwards \citep{lin1996}.  A second method is planet-planet scattering that puts exoplanets into eccentric orbits that are then tidally circularized \citep{chatterjee2008}. Another method is  Kozai migration, where a stellar companion on an inclined orbit induces exoplanets into eccentric orbits.  The primary star will then tidally circularize the planet \citep{wu2003}.   \citet{nagasawa2008} suggested that planet-planet scattering could lead to eccentric inclined exoplanets, which would then start Kozai cycles with the other exoplanets in the system resulting in hot Jupiters.  \citet{wu2011} proposed that secular chaos could cause the innermost planet to lose angular momentum and then be tidally circularized forming a hot Jupiter.  It is quite probable that all of these methods occur in nature, but we do not know which is the dominant mechanism. 

The presence of a stellar companion to HD 2638 does not help us differentiate between these methods.  While Kozai migration requires a third body in the system, the other methods are not ruled out by the existence of a stellar companion. To gain a deeper understanding of the dynamical evolution of this system we will need to monitor the system to eventually compute an orbit of the stellar system. Also continued RV monitoring is needed to identify any additional planets in long periods or highly inclined orbits.  Continued monitoring would allow the measurement of the RV signature of the companion star which will allow a more complete orbit to be determined.  

It is conceivable that the newly discovered component, C, is a close binary  masquerading as an exo-planet.  We estimate that C is fainter than B by about 4 mag in the visible, so the observed RV amplitude of 67.4\,m~s$^{-1}$  in the combined light of B+C corresponds to the 2.7\,km~s$^{-1}$   amplitude in the velocity of C. The minimum mass of the hypothetic binary companion Cb is then 0.012 solar, so it could be a 0.1 ${\cal M}_\odot$ dwarf in a 3.44-day orbit with a $30^\circ$ inclination.  This is possible, although statistically unlikely (probability 0.13). If, on the other hand, Ca and Cb had comparable masses, the  RV amplitude of this putative binary would be about 70\,km~s$^{-1}$,  and an orbital inclination of $\sim 20^\circ$ would suffice for blending with the lines of B and producing the fake exo-planet RV signal.  In the light of the discovery of C, a more in-depth analysis of the spectra of HD 2638 is desirable to validate the exo-planet. 



\section{30 ARI B}\label{30AriB}
  
30 Ari B (HIP 12184 = HD 16232 = WDS 02370+2439) is a   star classified as F6V  \citep{guenther2009} at a distance of 40.8$\pm$  pc \citep{vanLeeuwen2007}. Along with HIP 12189, it forms the  binary system 30 Ari. The pair has  a separation of $\sim$40\arcsec~and an estimated period of 34,000 years \citep{tokovinin2006}.  \citet{shaya2011} determined that the pair had a near 100\% probability of being a physical system. 30 Ari A (HIP 12189 = HD 16246)  has been listed as an F6III, but \citet{guenther2009} analyzed its colors and concluded it is actually an F5V star.  30 Ari A is also a spectroscopic binary with a period of 1.109526\,d in a nearly circular orbit \citep{morbey1974}.  30 Ari A has a measured age of 0.86$\pm$0.63 Gyr and 30 Ari B has a measured age of  0.91$\pm$0.83 Gyr \citep{guenther2009}.   Both stars are slightly metal rich \citep{guenther2009}.  The  details of the system are also given in Table \ref{tab:objects}.

Using the RV technique, \citet{guenther2009} discovered a 9.88$\pm$0.94 M$_{Jup}$ exoplanet orbiting 30~Ari B with a period of 335.1$\pm$2.5 d and an eccentricity of 0.289$\pm$0.092.  Based on the star's measured rotational velocity and the average rotational velocity of F-type stars, \citet{guenther2009} concluded that the star is being observed  nearly equator on. Assuming that the planet's orbital axis aligns with the star's spin axis, the minimum mass is close to the true mass. With just over six years of data, there are no indications of additional signals in the RV data coming from additional companions.
 
\begin{figure*}[htb]
    \centering
\includegraphics[height=6cm]{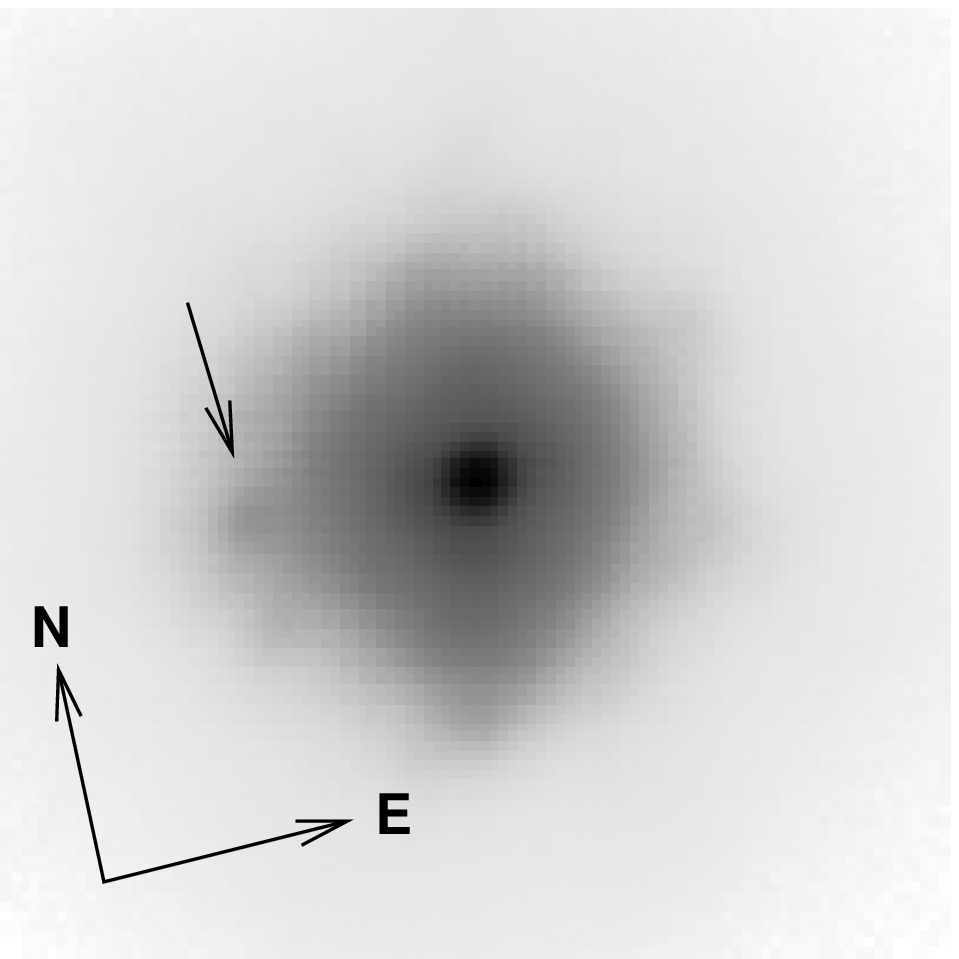}\label{subfig_A}  
\includegraphics[height=6cm]{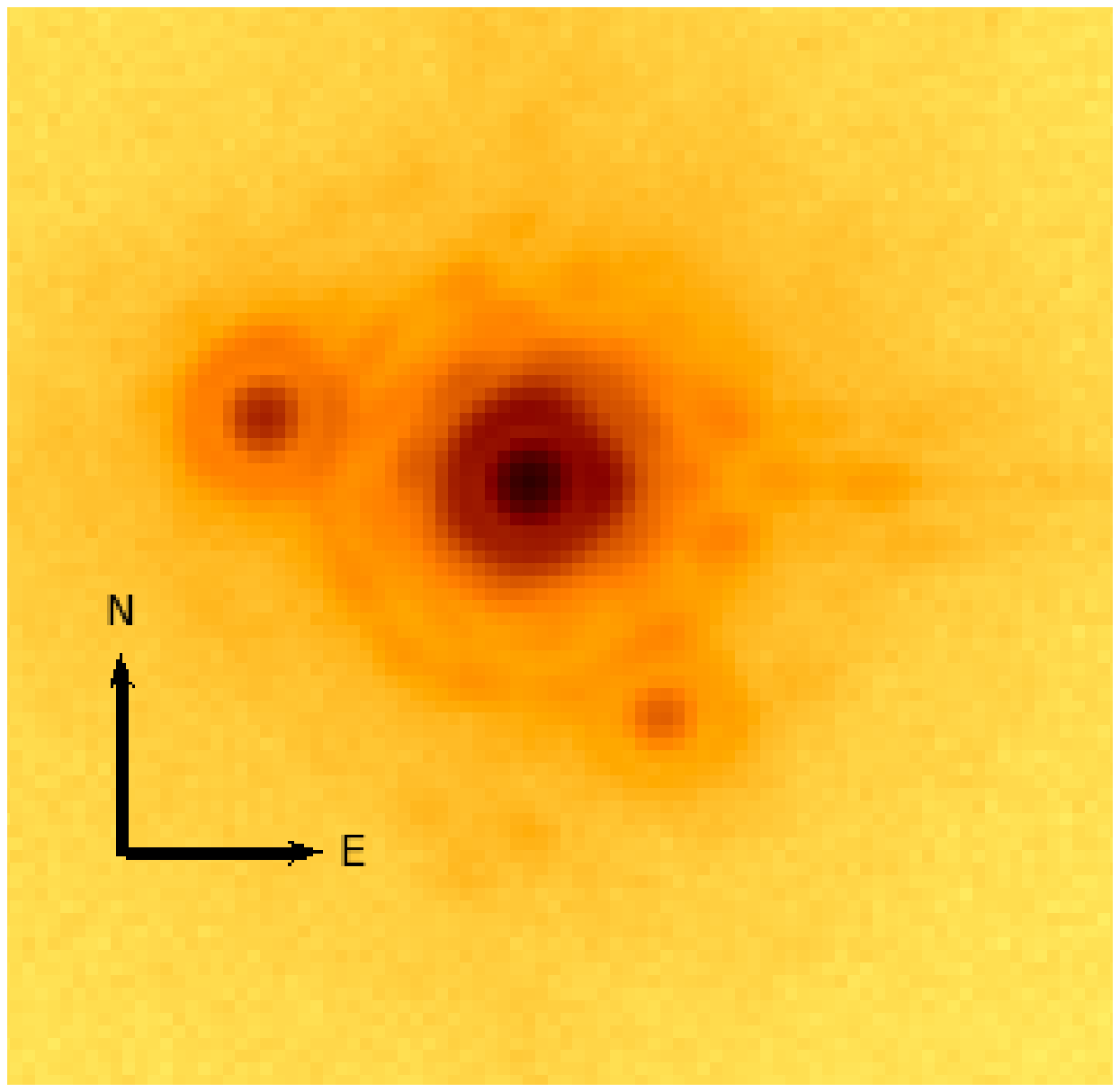}\label{subfig_B}
    \caption{ On the left is an $i$ image of the 30 Ari BC from Robo-AO.  It is a subimage of the full 44\arcsec~field of view and is $\sim$2\farcs5 across. The location of the newly discovered companion is indicated by the arrow. On the right is a $Ks$ image of the 30 Ari BC acquired with the PALM 3000 AO system. The object to the lower right of the primary star of the image is a ghost image caused by the neutral density filter.  This is a subimage of the full 25\arcsec~field of view and is $\sim$2\arcsec~across.     }
    \label{30Ari_image}
\end{figure*}

\citet{riddle2015} detected a companion to 30 Ari B with Robo-AO on 2012 September 3 UT at a separation of just over half an arcsecond.  The compainion was confirmed with Robo-AO approximately a year later and we conducted follow up observations in the near-IR with the Palm 3000 AO system on the Palomar 5 m telescope at approximately the same time.  Images of 30 Ari BC are shown in Figure \ref{30Ari_image}.   The measured astrometry and photometry are in  Table \ref{results_30ari}.   The error bars of 30 Ari BC are smaller than the error bars of HD 2638 BC because 30 Ari BC is a brighter star and AO performance is improved on brighter targets. 

30 Ari B has a proper motion of 0.151 $\pm $0.00075\arcsec/yr \citep{vanLeeuwen2007}.   Between the first and last observation  the companion moved 0.0229 $\pm$ 0.0199\arcsec/yr.  From this we conclude that the companion has common proper motion with 30 Ari B.     Like HD 2638, this would normally be designated 30 Ari Bb, but to avoid confusion with the planet we designate it as 30 Ari C.   The measured motion is consistent with orbital motion in that both the position angle and separation are changing in a regular manner.  With only three data points it is not possible to determine if there is a deviation from linear motion, which would be required to confirm orbital motion. 

\begin{figure*}[htb]
   \centering
   \includegraphics[height=60mm]{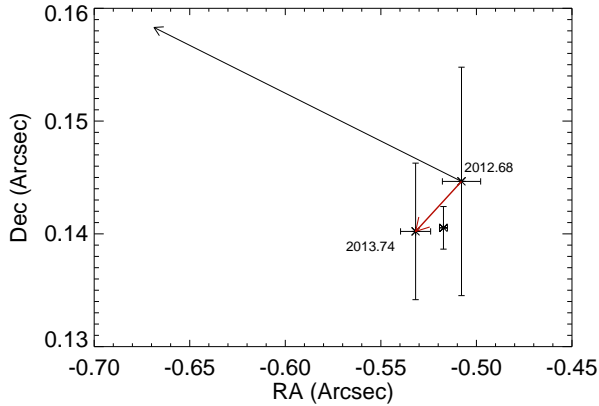}  
   \caption{ The measured astrometry for 30 Ari BC is plotted. The black arrow is the position the companion would have been at our last observation if it was a background star with no proper motion.  The red line shows the actual motion between our first and last observation.  }
   \label{30AriB_motion}
\end{figure*}

As part of a duplicity survey of exoplanet host stars, \citet{ginski2012} observed 30 Ari B with  the 2.2\,m Calar Alto telescope and the AstraLux lucky imaging camera in the SDSS \textit{i} filter on 23 February 2010.  They did not detect the companion due to the limited dynamic range of the lucky imaging system and poor weather.  

\subsection{DISCUSSION}

The projected separation between B and C is 22.3\,AU and corresponds to the statistical period $P^* = 80$\,yr. Using the same technique as used for HD~2638, we computed photometry and masses for 30 Ari BC which are presented in Table \ref{tab:ptm}.  The luminosities and masses of B and C correspond to spectral types F8V and M1V, respectively. 

As in the case of HD~2638~B, the possibility that the exoplanet signal comes from a hierarchical multiple system needs to be addressed.  According to \citep{guenther2009}, the axial rotation of B is 38\,km~s$^{-1}$, while the RV amplitude is 0.27\,km~s$^{-1}$. Although  at visible wavelengths C is fainter than B by at least 5 mag, it has a lower effective temperature and stronger lines, making it difficult to evaluate the amplitude of the hypothetic binary Ca,Cb required to mimick the exo-planet. An amplitude of  10\,km~s$^{-1}$, for example, can be produced by a companion Cb of  0.26 ${\cal M}_\odot$ at $90^\circ$ inclinantion.   
 
\citet{guenther2009}  points out that, ``...according to \citet{eggenberger2007} most of the very massive planets are in binary systems. But ... the distance to 30 Ari A is presumably too large to have any effect on the properties of the planet.''  Our discovery of 30 Ari C may explain why it is not the exception.  

The detection of a stellar companion to 30 Ari B means that the 30 Ari system is a quadruple system containing at least one exoplanet.  This makes it the second confirmed quadruple system known to host an exoplanet. The first is  KIC 4862625 hosting the planet Ph1b \citep{schwamb2013}.  Ph1b is a circumbinary planet orbiting an eclipsing binary consisting of G- and M-type dwarfs.  The other  two components of the quadruple are in a tight visual binary  with a projected separation of 1000AU from the G/M pair.

The similarity between KIC 4862625 and 30 Ari is that both systems are quadruples consisting of two relatively close pairs that are widely separated. These 2+2 systems are relatively common. \citet{riddle2015} found that over 10\% of wide binaries are actually 2+2 quadruples and that the majority of quadruples are in this form, rather than the form of a triple system with a distant companion (i.e. 3+1).  In both of the quadruple systems hosting exoplanets, the two binaries making up the quadruple are most likely far enough apart that there is no impact on the dynamics or the formation of the exoplanet.  

\begin{deluxetable}{lllcccc}
\tablewidth{0pt}
\tablecaption{Relative astrometry and photometry of 30 Ari B. \label{results_30ari}}
\tablehead{ \colhead{Epoch} & \colhead{$\theta$ ($^{\circ}$)} &\colhead{$\rho$ (\arcsec)} & \colhead{\textit{$\Delta$i}} & \colhead{\textit{$\Delta$J}}&\colhead{\textit{$\Delta$Ks}}& \colhead{Instrument} }
\startdata
2012.6750 & 285.9\phn$\pm$1.1 & 0.528$\pm$0.010   & 4.32$\pm$0.1 & ...           & ...           & Robo-AO\\
2013.6167 & 285.2\phn$\pm$0.2 & 0.536$\pm$0.002   & 4.31$\pm$0.1 & ...           & ...           & Robo-AO\\
2013.7426 & 284.77$\pm$0.79   & 0.55\phn$\pm$0.008 & ...          & 3.44$\pm$0.15 & 2.84$\pm$0.07 & PHARO\\
\enddata
\end{deluxetable}


\section{SUMMARY}\label{summary}

We have detected stellar companions to the exoplanet host stars HD 2638 and 30 Ari B.  Both systems are part of wider multiple star systems. In the case of 30 Ari B, the other component, 30 Ari A, is also a binary star making it a quadruple system.  This is the second quadruple system known to host an exoplanet.  

\citet{bonavita2007} found that binaries with exoplanets tend to have low mass stellar companions.  Both of the systems detected here continue that trend. This could be a selection effect, as most RV surveys have eliminated known binaries from their observing lists. These known binaries were mostly discovered by the  Hipparcos satellite, visual micrometers or speckle interferometry instruments, which all have limited dynamic range.  AO is currently being used for most exoplanet duplicity searches and it can achieve a much higher dynamic range allowing for the detection of late-type stars.  

HD 2638 BC and 30 Ari BC both have projected separations of less than 30 AU and add to the rather small number of systems with such low separations.    The stellar companions have the potential to have been a key part of the dynamical evolution of the exoplanets.  However, the veracity of these exoplanets has to be addressed in light of the new companions, to eliminate the possibility that these RV discoveries are false positives caused by hierarchical multiplicity.    These systems strengthen the link between massive planets and binary stars. The astrometry of these binaries should continued to be monitored.  In the near future this will hopefully detect orbital motion and provide an indication to the period of the objects.  With a long enough span of observations, an orbit can be computed, which will provide a more complete picture of the nature of the dynamical interaction between the binary companion and the exoplanet. In addition to the system's astrometry, their radial velocities should also be monitored. This may reveal a trend in the RV caused by the stellar companion.  Knowledge of the RV trend coupled with measured astrometry will enable the computation of the full orbital solution. 


\acknowledgements

We thank the staff of the Palomar Observatory for their invaluable assistance in collecting these data. This paper was based in part on observations obtained at the Hale Telescope, Palomar Observatory as part of a continuing collaboration between the California Institute of Technology, NASA/JPL, Oxford University, Yale University, and the National Astronomical Observatories of China. A portion of the research in this paper was carried out at the Jet Propulsion Laboratory, California Institute of Technology, under a contract with the National Aeronautics and Space Administration (NASA). The Robo-AO system is supported by collaborating partner institutions, the California Institute of Technology and the Inter-University Centre for Astronomy and Astrophysics, by the National Science Foundation under Grant Nos. AST-0906060,  AST-0960343, and AST-1207891, by a grant from the Mt. Cuba Astronomical Foundation and by a gift from Samuel Oschin.  C.B. acknowledges support from the Alfred P. Sloan Foundation. This research made use of the Washington Double Star Catalog maintained at the U.S. Naval Observatory, the SIMBAD database, operated by the CDS in Strasbourg, France and NASA's Astrophysics Data System.   

{\it Facilities:} \facility{ \facility{PO:1.5m (Robo-AO), Hale (PHARO)} } 
 


\end{document}